\documentclass[12pt]{spieman}  
\usepackage{amsmath,amsfonts,amssymb}

\DeclareMathOperator*{\argmin}{arg\,min}
\usepackage{graphicx}
\usepackage{setspace}
\usepackage{tocloft}
\usepackage{comment}
\usepackage{lineno}

\title{Implicit Electric Field Conjugation with a Photonic Lantern Nuller}

\author[a]{Yinzi Xin} 
\author[a]{Daniel Echeverri} 
\author[a]{Nemanja Jovanovic} 
\author[b]{Jonathan Lin} 
\author[b]{Yoo Jung Kim} 
\author[a,c]{Dimitri Mawet} 
\author[d]{Sergio Leon-Saval} 
\author[e]{Rodrigo Amezcua-Correa}
\author[e]{Stephanos Yerolatsitis}
\author[b]{Michael P. Fitzgerald} 
\author[a]{Pradip Gatkine} 
\author[a,f]{Suvinay Goyal} 
\author[g]{Barnaby Norris} 
\author[c]{Garreth Ruane} 
\author[h]{Steph Sallum}

\affil[a]{Department of Astronomy, California Institute of Technology, 1200 E California Blvd, Pasadena, CA, 91125, USA}
\affil[b]{Department of Physics \& Astronomy, 430 Portola Plaza, University of California, Los Angeles, CA 90095, USA}
\affil[c]{Jet Propulsion Laboratory, California Institute of Technology, 4800 Oak Grove Drive, Pasadena, CA, 91109, USA}
\affil[d]{Sydney Astrophotonic Instrumentation Laboratory, School of Physics, The University of Sydney, Sydney, NSW 2006, Australia}
\affil[e]{The College of Optics and Photonics, University of Central Florida, 4304 Scorpius St, Orlando, FL 32816}
\affil[f]{University of Illinois Urbana-Champaign, Department of Astronomy, Urbana, Illinois, USA}
\affil[g]{Sydney Institute for Astronomy, School of Physics, Physics Road, The University of Sydney, NSW 2006, Australia}
\affil[h]{Department of Physics \& Astronomy, University of California, Irvine, 4129 Frederick Reines Hall, Irvine, CA 92697 USA}

\cftpagenumbersoff{figure}
\cftpagenumbersoff{table} 
\begin{document} 
\maketitle

\begin{abstract}
The Photonic Lantern Nuller (PLN) is an instrument concept designed to characterize exoplanets within a single beam-width from its host star. The PLN leverages the spatial symmetry of a mode-selective photonic lantern (MSPL) to create nulled ports, which cancel out on-axis starlight but allow off-axis exoplanet light to couple. The null-depths are limited by wavefront aberrations in the system as well as by imperfections in the lantern. We show that the implicit electric field conjugation algorithm can be used to reduce the stellar coupling through the PLN by orders of magnitude while maintaining the majority of the off-axis light, leading to deeper null depths ($\sim 10^{-4}$) and thus higher sensitivity to potential planet signals. We discuss a theory for the tradeoff we observed between the different ports, where iEFC improves the nulls of some ports at the expense of others, and show that targeting one port alone can lead to deeper starlight rejection through that port than when targeting all ports at once. We also observe different levels of stability depending on the port and discuss the implications for practically implementing this technique for science observations.
\end{abstract}


{\noindent \footnotesize\textbf{*}Yinzi Xin,  \linkable{yxin@caltech.edu} }

\begin{spacing}{2}   

\section{Introduction}
\label{sect:intro}
The characterization of exoplanets was identified by the Decadal Survey for Astronomy and Astrophysics 2020 as one of the top scientific priorities \cite{NRC_2020Decadal}. High-resolution spectroscopy is especially critical for many measurements, including that of the planet's radial velocity, spin, atmospheric composition, and surface features through Doppler imaging \cite{wang_hdc1}. It can also enable the potential detection of exomoons \cite{ruffio_exomoons}. The Photonic Lantern Nuller \cite{xin_2022, Tuthill2022-NIH} is an instrument concept that enables the detection and characterization of exoplanets at and within $1 \,\ \lambda/D$, where $\lambda$ is the wavelength and $D$ the telescope diameter. It is inspired by the Vortex Fiber Nuller (VFN) \cite{Ruane2018_VFN,echeverri_2019}, but unlike the VFN, which has only one nulled channel with a circularly symmetric coupling profile, the PLN provides four nulled channels, each with a unique coupling profile. This allows for more planet flux to be retained, and also helps place better constraints on the planet's flux ratio and spatial position (with a $180^{\circ}$ degeneracy in the position) \cite{xin_2022}. This also allows the PLN to be used for spectroastrometry \cite{kim_2024}, albeit with the same $180^{\circ}$ degeneracy as for planet localization.

The PLN exploits the symmetries of the ports of a six port mode-selective photonic lantern (MSPL) \cite{LeonSaval_MSPL}, a special type of photonic lantern \cite{LeonSaval_PL_2013} that utilizes dissimilar cores that enable ports to be mapped into linearly polarized (LP) modes, or the eigenmodes of a radially symmetric, weakly guiding step-index waveguide. Each mode at the multi-mode face (MMF) of the lantern is mapped to a single-mode fiber (SMF) output, such that light coupling to a given mode at the MMF side will result in flux in the corresponding SMF core. The symmetries of the modes corresponding to the MSPL ports results in a null at the center (to which the star is aligned), with finite transmission off-axis where potential planets may exist.

\begin{figure*}[t]
\begin{center}
	\includegraphics[scale = 0.45]{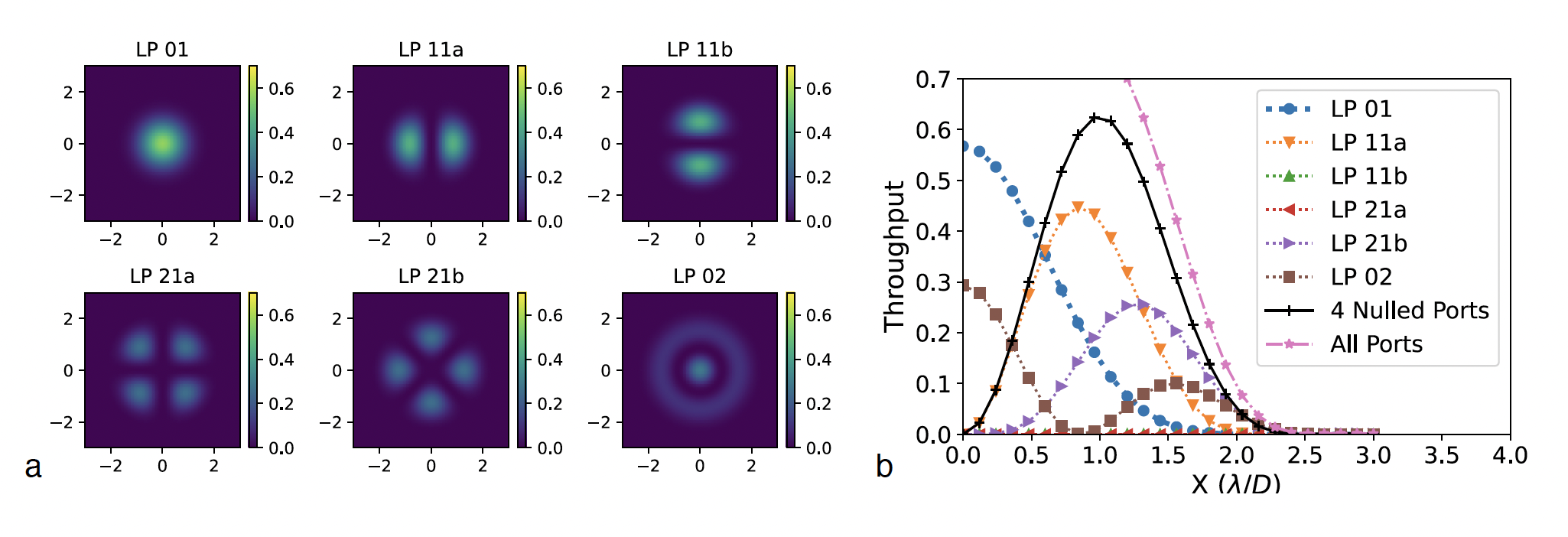}
	\caption{\label{fig:coupling_maps} a) Throughput maps for each port of an ideal six-port mode-selective photonic lantern, spanning from -3 $\lambda/D$ to 3 $\lambda/D$ in each direction. b) X-axis line profiles from the throughput maps. The four nulled ports are LP 11ab and LP 21ab. The line profile of the summed throughput of the nulled ports is shown in black with cross-hair markers. Figure adapted from Ref. \citenum{xin_2022}.}
\end{center}
\end{figure*}

The operating principles of the PLN are fully derived in Ref. \citenum{xin_2022}, and the first laboratory demonstration of the PLN (in both monochromatic and broadband light) is presented in Ref. \citenum{xin_2024_pln_lab}. The theoretical spatial coupling profiles from Ref. \citenum{xin_2022} are displayed in Fig. \ref{fig:coupling_maps}a, showing that four of the six ports of a mode-selective lantern are nulled as they have no coupling at the central location of the star. The X-axis cross-sections of these coupling maps are displayed in Fig. \ref{fig:coupling_maps}b, showing a theoretical planet throughput of over $60\%$ when summed across the nulled ports. In this work, we present the results of using wavefront control --- specifically the implicit electric field conjugation algorithm (iEFC) \cite{haffert_2023_iefc} --- to deepen the central nulls of the PLN.

\section{Implicit Electric Field Conjugation}

The implicit electric field conjugation algorithm for active suppression of starlight is described in Ref \citenum{haffert_2023_iefc}. We present a simplified overview of it here.

The stellar electric field can be modulated by applying probes on the deformable mirror (DM), and the electric field is linearly related to the difference between an image with some probe and the image with the same probe but with opposite sign. Minimizing the measurement $\delta$ --- a series of such `differenced' images --- thus also minimizes the electric field. We can empirically calibrate the influence of the DM on $\delta$ by applying a mode on the DM and encoding the change in $\delta$ that it produces into a response matrix:

\begin{equation}
    \delta = S \alpha,
\end{equation}

\noindent where $\alpha$ is the DM command and $S$ the calibrated response matrix. The basis set of modes whose influence is sequentially measured and encoded into $S$ are referred to as the ``calibration modes."

The iEFC solution for the desired DM input is given by

\begin{equation}
    \alpha = \argmin_\alpha |\delta+S\alpha|^2+\lambda|\alpha|^2 = -(S^TS+\lambda I)^{-1}S^T\delta = -C\delta.
\end{equation}

The parameter $\lambda$ can be set to penalize large DM solutions, and the control matrix $C=(S^TS+\lambda I)^{-1}S^T$ is computed ahead of time, after calibrations are complete.

Unlike the alternative electric field conjugation (EFC) algorithm \cite{giveon_efc}, which is model-based, iEFC is data-driven. It has the advantage of not being limited by model fidelity but also the disadvantage of requiring testbed calibration time. We choose to use iEFC, however, because it is relatively more advantageous for the PLN than for conventional coronagraphs, as the PLN has a much more limited field-of-view (FOV) and thus requires significantly fewer modes to be calibrated than for a coronagraphic dark hole. The number of calibrated modes required scales linearly with the area of the FOV, which scales quadratically with FOV radius, so the iEFC calibration overhead for the PLN (with a FOV radius of $\sim 2 \lambda/D$) compared to a typical coronagraphic dark hole (extending to $\sim 10 \lambda/D$) is smaller by a factor of $2^2/10^2$, or about $0.04$.

For this work, we use the implementation of iEFC from the \textsf{lina} package \cite{milani_lina}.

\section{Experimental Setup}

A detailed schematic of the front-end of Polychromatic Reflective Testbed (PoRT) can be found in Ref. \citenum{xin_2024_pln_lab}, along with the PLN coupling maps measured using the testbed without wavefront control --- i.e. the DM merely flattened by maximizing coupling through a single-mode fiber. The results obtained in Ref. \citenum{xin_2024_pln_lab} were obtained using a photodiode with only one input, and therefore had to be measured sequentially. However, for wavefront control, it is convenient to be able to measure the coupling through all the relevant ports simultaneously.

\begin{figure*}[t]
\begin{center}
	\includegraphics[scale = 0.45]{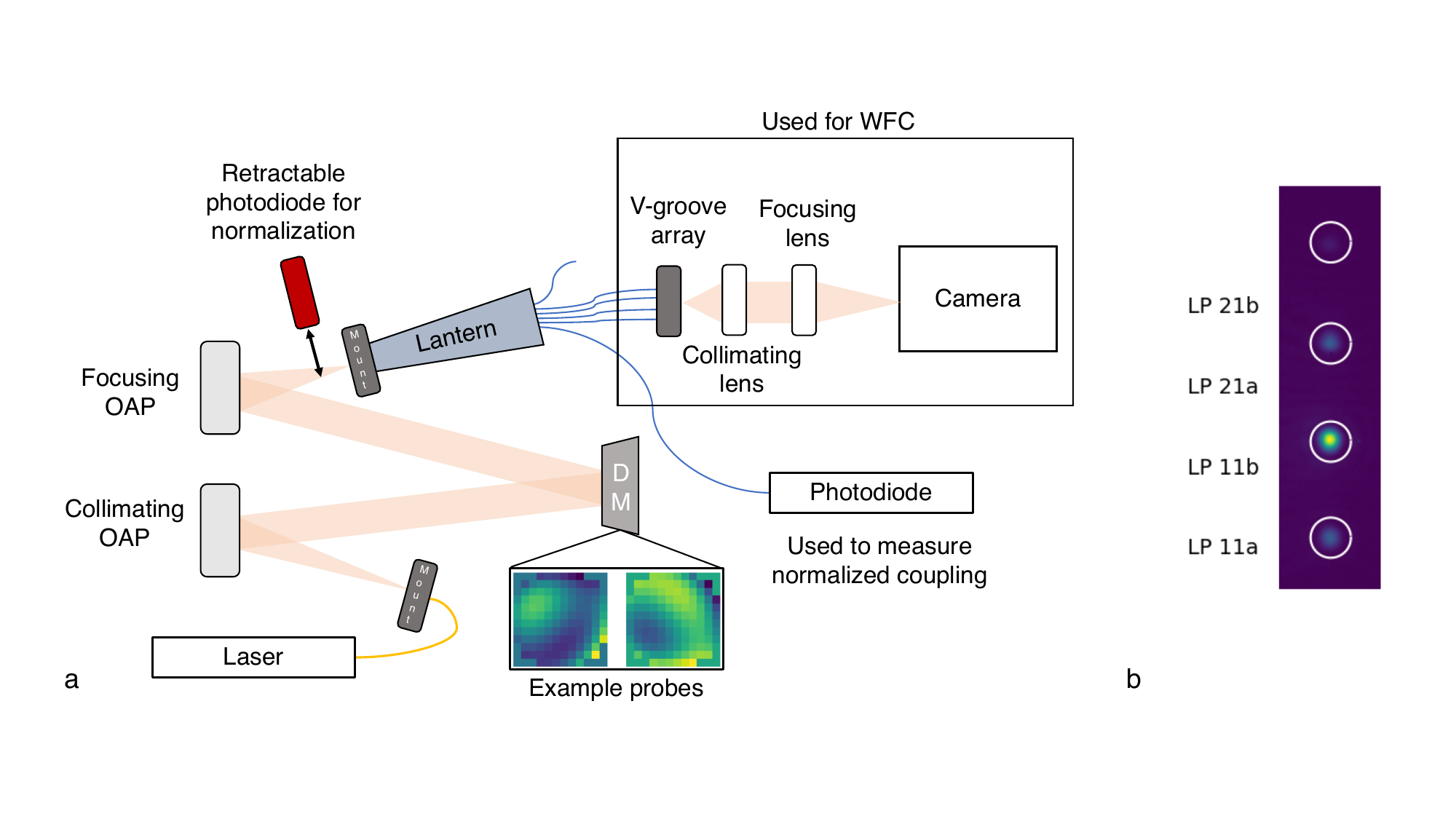}
	\caption{\label{fig:experimental_setup} a) A simplified diagram of the experimental setup. A monochromatic 1568.772 nm laser is injected into the bench, and the beam is collimated. A $12 \times 12$ deformable mirror can be used to manipulate the wavefront of the beam before it is focused onto the lantern ; the inset shows example DM probe modes $p_{1_+}$ and $p_{2_+}$. The SMF outputs of the lantern can then be routed to either a V-groove array to be imaged onto the camera, or to the photodiode. The photodiode is calibrated to a photometer that can slide into the beam just before the lantern, and thus provides measurements normalized to the incoming beam. While performing wavefront control, all four nulled ports are routed to the V-groove, with the non-nulled ports disconnected in order to not saturate the camera. After performing wavefront control, we sequentially route the nulled outputs to the coupling photodiode to obtain coupling maps for each port. The coupling photodiode is calibrated to a photodiode that can be inserted before the lantern in order to normalize the coupling maps to the incident light on the lantern, providing normalized measurements of $\eta$. b) The image on the camera cropped to the region of interest. The white circles indicate the photometric apertures used to measure coupling through the lantern's nulled ports.}
\end{center}
\end{figure*}

In Figure \ref{fig:experimental_setup}a, we present an updated simplified diagram of PoRT, which now includes a back-end that images the outputs of the nulled ports onto a camera, allowing the fluxes coupled into each port to be simultaneously measured using photometry. As in Ref. \citenum{xin_2024_pln_lab}, we use a Thorlabs TLX2 laser set to a wavelength of 1568.772 nm injected into the bench with a polarization maintaining fiber, and use a tunable iris to set the $F\#$ of the beam being injected into the lantern to maximize the throughput into the LP 11 ports, resulting in $F\#=6.2$. Note that one polarization of light is used for this experiment. Ref. \citenum{xin_2024_pln_lab} showed that the lantern exhibited polarized differences at the level of a few $10^{-3}$ (in coupled intensity) between two orthogonal polarizations of light. With conventional coronagraphs, these polarized differences would limit the achievable contrast due to different DM solutions being needed for each polarization \cite{ashcraft_2025}. However, the overlap integral of the PLN allows for more nulling degrees of freedom than conventional coronagraphs with a focal-plane detector (see Sec. \ref{sec:disc}). This potentially enables joint solutions that null both polarizations simultaneously despite the differences in mode shape, a study that is left for future work.

Figure \ref{fig:experimental_setup}b shows an example camera image, cropped to the region containing the lantern outputs and with the dark frame (the camera image when the light source is turned off) subtracted. The overlaid circles indicate the apertures used for photometry, where each intensity measurement is the sum of the counts contained within the defined circle. When the incoming beam is aligned to the `center' of the lantern (which we choose as the location of minimum summed coupling through the nulled ports), these intensity measurements correspond to the stellar coupling through each port of the lantern. These intensity measurements are proportional to $\eta_s$, or the fraction of the incoming starlight coupled into each port. For closed-loop wavefront control, all four nulled ports are routed to the V-groove, and we work directly with the intensity measurements made on the camera (i.e. the sum of the counts within each defined aperture). We leave the non-nulled ports disconnected in order to not saturate the camera. After performing wavefront control, we sequentially route the nulled outputs to the coupling photodiode to obtain coupling maps for each port. The coupling photodiode is calibrated to a photodiode that can be inserted before the lantern in order to normalize the coupling maps to the incident light on the lantern, providing normalized measurements of $\eta$.

\section{Implementation and Results}

\subsection{Example targeting all nulled ports} \label{sec:four_ports}
We define the following two sets of probes, where $Z_n$ is the $n$th Noll-ordered Zernike mode defined across the $12 \times 12$ DM actuator grid:

\begin{align}
    p_{1_{\pm}} &= \pm (Z_5+Z_6+Z_7+Z_8)/\sqrt{4} \\
    p_{2_{\pm}} &= \pm (Z_5-Z_6+Z_7-Z_8)/\sqrt{4}
\end{align}

The probe mode shapes of $p_{1_+}$ and $p_{2_+}$ are shown in the inset of Fig. \ref{fig:experimental_setup}a. This choice of probe is motivated by the fact that the LP 11 modes are primarily sensitive to Coma ($Z=5$ and 6) while the LP 21 modes are primarily sensitive to Astigmatism ($Z=7$ and 8). In fact, simulations show that the dominant modes of $S$ when sensed using \textit{completely random} probes that modulate the whole DM, as obtained through a singular-value decomposition (SVD), are indeed Coma and Astigmatism first, followed by higher-order irregular modes. We find that in the presence of noise, however, using $p_1$ and $p_2$ as defined results in much better signal and iEFC performance than using more complex probes.

\begin{figure*}[t]
\begin{center}
	\includegraphics[scale = 0.45]{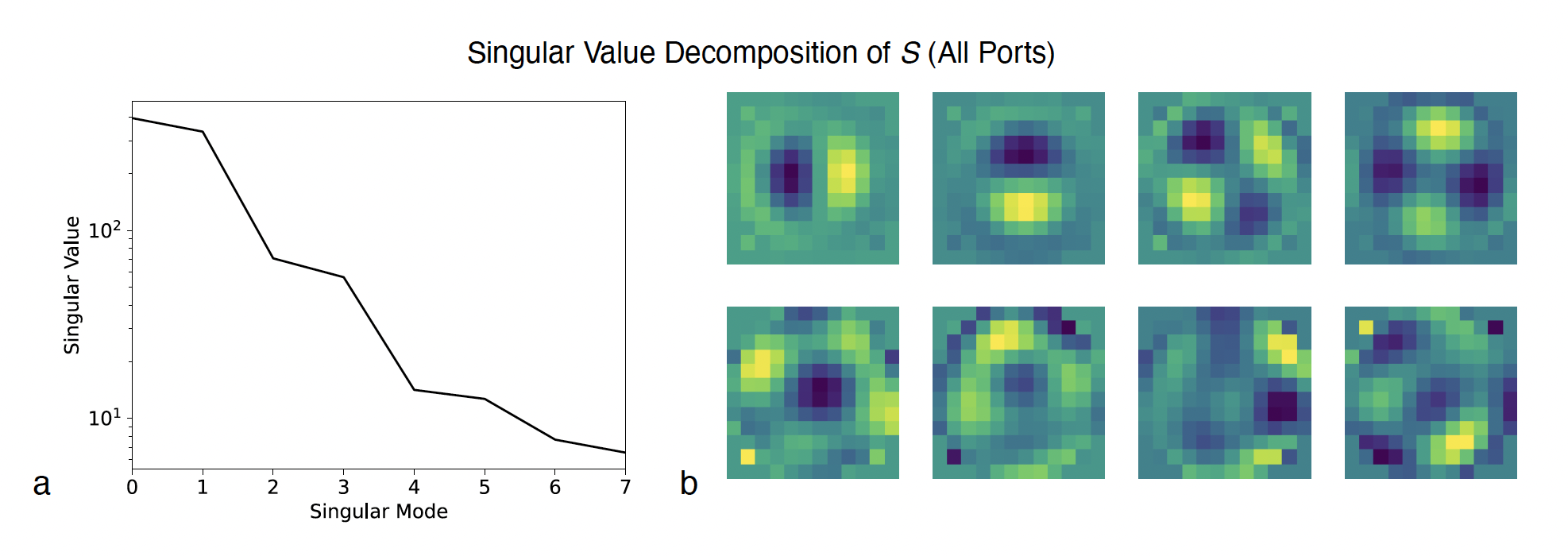}
	\caption{\label{fig:S_svd} a) The singular values of the response matrix $S$ when measuring all four nulled ports. b) The corresponding singular modes in DM space, arranged in order of descending singular value.}
\end{center}
\end{figure*}

Using a probe amplitude of 0.02 and a calibration mode (i.e. the basis vectors for representing DM inputs) amplitude of 0.01 (both in DM control units that range from 0 to 1, mapping to the control voltage range of the DM, which has a 3.5 $\mu$m stroke), we then calibrate the response matrix $S$ across the set of Zernikes modes from $Z_4$ to $Z_{30}$ (a total of 27 modes). The singular values and singular modes of $S$ in DM space (arranged in order of descending singular value) are displayed in Fig. \ref{fig:S_svd}, showing the dominance of Coma and Astigmatism in the instrumental response. From $S$, we calculate the control matrix $C$ using $\lambda$ equal to 1/10 of the maximum diagonal value of $S^TS$.

\begin{figure*}[t]
\begin{center}
	\includegraphics[scale = 0.45]{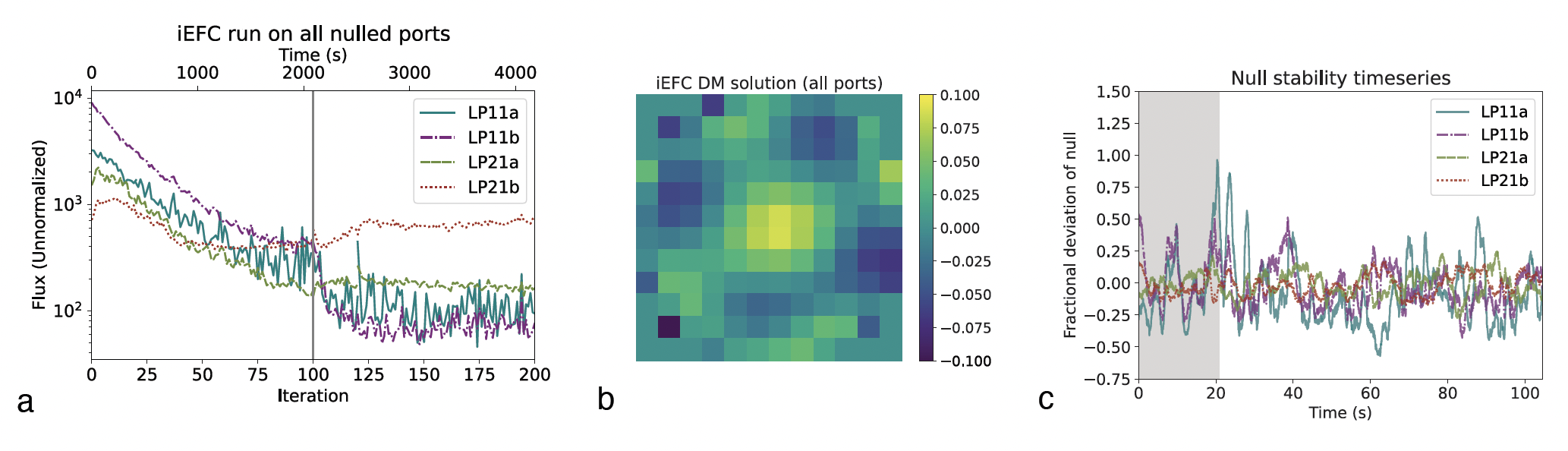}
	\caption{\label{fig:all_ports_run} a) An example iEFC run targeting all four nulled ports. The vertical gray line indicates a change in camera integration time along with a control matrix recalibration. The final null depths from this run are compared to the initial null depths in Table \ref{tab:metrics}. b) The DM solution found by iEFC targeting all four nulled ports at once. The root-mean-square (RMS) of the DM control values is 0.033, corresponding to approximately 120 nm of stroke. c) A timeseries of the fractional fluctuations in the null obtained with iEFC, spanning the timescale of one iEFC iteration. The gray shaded region corresponds to the timescale of one photometric measurement, and each iEFC iteration takes five photometric measurements --- four for the probes and one without any probe. The LP 11a and LP 11b nulls exhibit significant fluctuations over the timescale of an iEFC iteration, showing that these nulls are likely limited by testbed instability and not the lantern itself. }
\end{center}
\end{figure*}

\begin{figure*}[t]
\begin{center}
	\includegraphics[scale = 0.6]{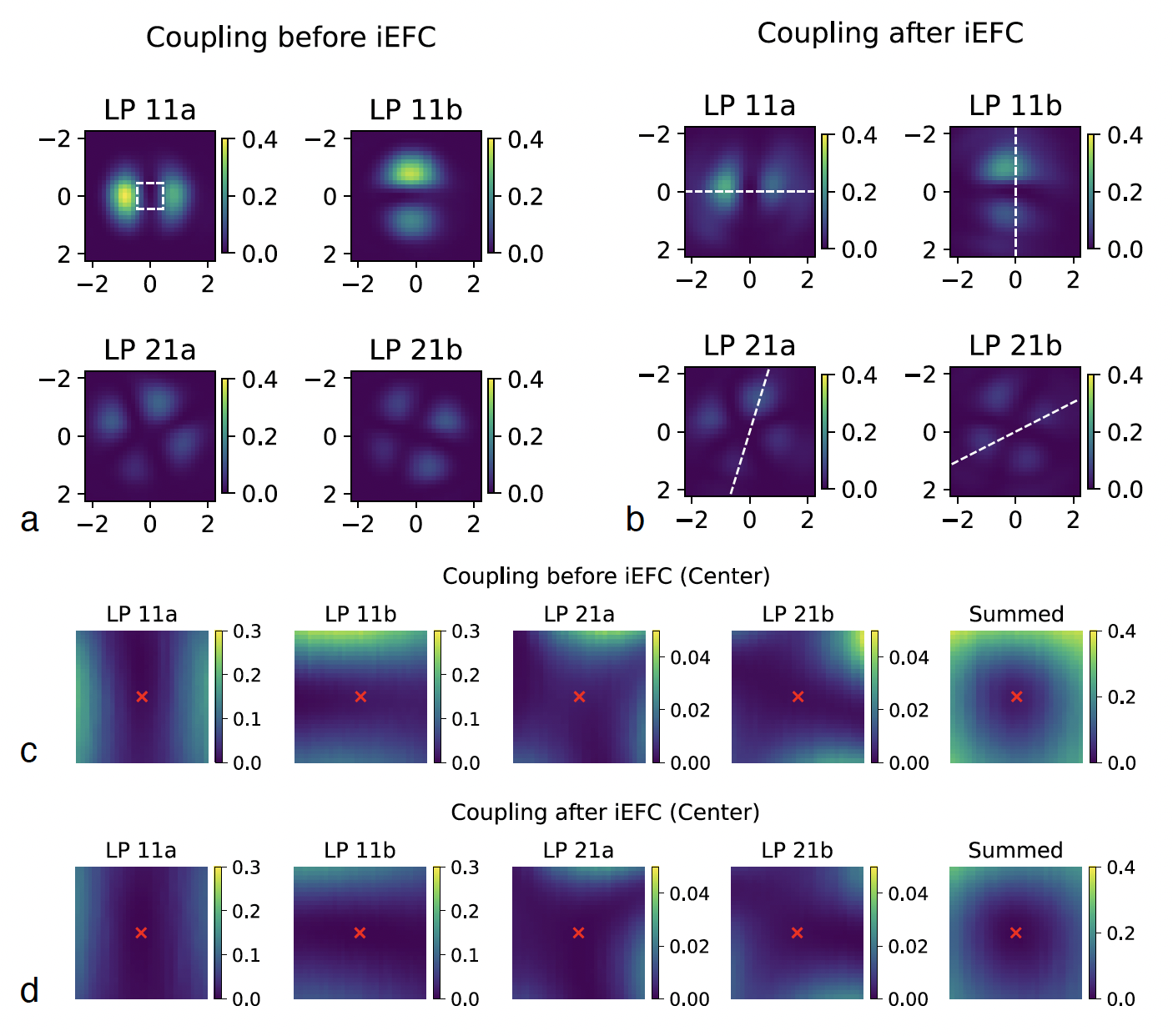}
	\caption{\label{fig:all_ports_maps} a) The normalized coupling maps through the nulled ports with the original DM surface, with spatial extent in units of $\lambda/D$. The white dashed box in the first panel indicates the finely-sampled region shown in (c) and (d) b) The normalized coupling maps using the DM solution found with iEFC, with spatial extent in units of $\lambda/D$. The dashed white lines indicate the locations of the S/N ratio cross-sections shown in Fig. \ref{fig:all_ports_crosssec}. c) Finely-sampled coupling maps of the lantern center with the original DM surface, spanning 1/5 of the spatial extent in part (a). d) Finely-sampled coupling maps of the lantern center using the DM solution found with iEFC, spanning 1/5 of the spatial extent in part (a). The red crosses indicate the location where the beam is aligned for the camera measurements, and also where $\eta_s$ is measured. We observe that iEFC spatially redistributes the coupling values through the lantern, lowering the coupling in the middle of the field of view and causing a diffuse extension of the coupling distribution at larger separations.}
\end{center}
\end{figure*}

\begin{figure*}[t]
\begin{center}
	\includegraphics[scale = 0.5]{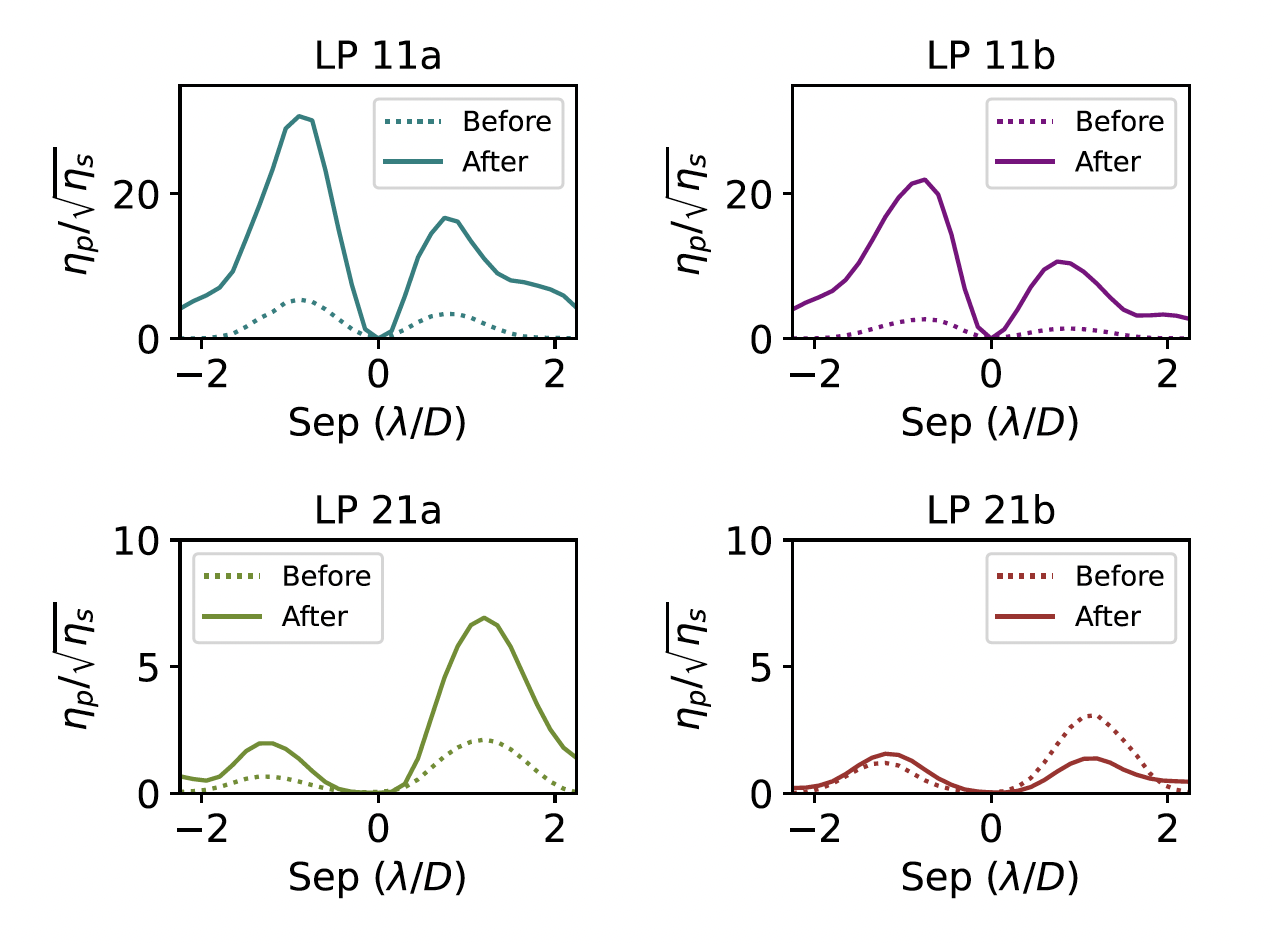}
	\caption{\label{fig:all_ports_crosssec} A comparison of the relative S/N ratio $\eta_p/\sqrt{\eta_s}$ along the cross-sections indicated in Fig. \ref{fig:all_ports_maps}b, before and after performing iEFC. In this case, wavefront control significantly improves the S/N ratio for three of ports while slightly degrading it in one. A theory for the observed tradeoff between ports is discussed in Section \ref{sec:disc}.}
\end{center}
\end{figure*}

Figure \ref{fig:all_ports_run}a shows an example closed-loop iEFC run, which significantly deepens the nulls in three of the ports, while the last null ends up approximately the same as it started. Normalized coupling maps are obtained using the photodiode, both with the original DM map and with the DM map after running iEFC, and are presented in Figure \ref{fig:all_ports_maps}. The metrics of stellar coupling, planet coupling, and null depth before and after iEFC are presented in Table \ref{tab:metrics}, and select cross-sections of the relative signal-to-noise (S/N) ratio, given by $\eta_p/\sqrt{\eta_s}$, before and after iEFC are also presented in Figure \ref{fig:all_ports_crosssec}.

\begin{table}[h]
\caption{The values of $\eta_{p_{peak}}$ (the peak planet coupling, or the maximum coupling value for each port) and $\eta_s$ (the stellar coupling) for the four nulled ports before and after performing iEFC, targeting all nulled ports at once. Also shown is the null depth ($\eta_s$/$\eta_{p_{peak}}$) for each port before and after performing iEFC.}

\begin{center} \label{tab:metrics}

\begin{tabular}{|c|c|c|c|c|} \hline 

& \textbf{LP 11a} & \textbf{LP 11b} & \textbf{LP 21a} & \textbf{LP 21b}  \\ \hline 

$\eta_{p_{\text{peak}}}$ (Before)& 0.394 & 0.358 & 0.131 & 0.112 \\ \hline 

$\eta_{p_{\text{peak}}}$ (After)& 0.246 & 0.224 & 0.108 & 0.0715 \\ \hline 

$\eta_s$ (Before)& $5.34 \times 10^{-3}$ & $1.65 \times 10^{-2}$ & $3.70 \times 10^{-3}$ & $1.31 \times 10^{-3}$ \\ \hline 

$\eta_s$ (After)& $5.80 \times 10^{-5}$ & $8.10 \times 10^{-5}$ & $2.35 \times 10^{-4}$ & $1.15 \times 10^{-3}$\\ \hline 

$\eta_s/\eta_{p_{\text{peak}}}$ (Before)& $1.35 \times 10^{-2}$ & $4.62 \times 10^{-2}$ & $2.83 \times 10^{-2}$ & $1.16 \times 10^{-2}$ \\ \hline 

$\eta_s/\eta_{p_{\text{peak}}}$ (After)& $2.36 \times 10^{-4}$ & $3.62 \times 10^{-4}$ & $2.18 \times 10^{-3}$ & $1.61 \times 10^{-2}$ \\ \hline 

\end{tabular}
\end{center}
\end{table}

This experiment shows a typical tradeoff that we observe with iEFC with the PLN, where certain nulls become degraded to deepen others if it reduces the stellar coupling overall. In this case, the LP 11 nulls have likely reached the limit imposed by the stability of the testbed, as measurements of the null with the loop open show significant fluctuations on the timescale of an iEFC measurement (Fig. \ref{fig:all_ports_run}c). Meanwhile, the LP 21 ports have not reached the limit imposed by testbed stability. This can be demonstrated by running iEFC on them individually as in Section \ref{sec:single_port}, which results in deeper nulls in those ports than when running iEFC on all four ports at once. In Section \ref{sec:disc}, we discuss simulations that provide more insight into this observed tradeoff, as well as the predicted behavior of iEFC under various conditions that are not realizable on the actual testbed.

\subsection{Example targeting a single port} \label{sec:single_port}
For these runs targeting a single port, LP 21b, we use a different set of probes that better captures the effect of the control modes on this port. Simulations show that while the most dominant mode of $S$ in this case is Astigmatism, as expected, the second dominant mode has a complex and irregular shape (somewhat resembling the eventual simulated DM solution), and thus sensing it well requires probes comprised of many Zernikes:

\begin{align}
    p_{3_{\pm}} &\propto  \pm \sum_{n=4}^{30} Z_n\\
    p_{4_{\pm}} &\propto \pm \sum_{n=4}^{30} (-1)^n Z_n.
\end{align}

In this case, $p_{3_{\pm}}$ and $p_{4_{\pm}}$ are each normalized to have an RMS of 1 before being multiplied by the desired probe amplitude in DM control units.

\begin{figure*}[t]
\begin{center}
	\includegraphics[scale = 0.5]{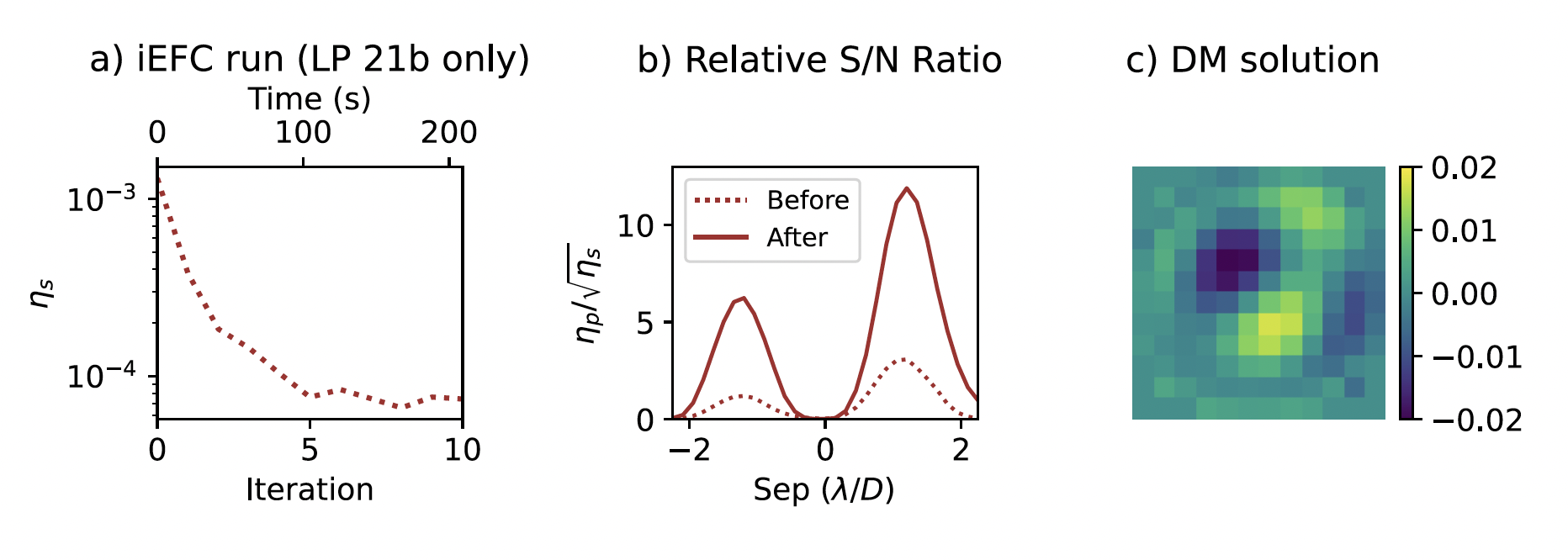}
	\caption{\label{fig:lp21b_only} a) An example iEFC run targeting only the LP 21b port, showing significant improvement in the null. The final $\eta_s = 7.42\times 10^{-5}$ is extracted from camera photometry and converted to a normalized coupling value based on the calibrated initial measurement. b) The relative S/N ratio ($\eta_p/\sqrt{\eta_s}$) along the cross-section indicated in Fig. \ref{fig:all_ports_maps}b, before and after running iEFC. The peak coupling $\eta_{p_{peak}}$ is 0.110, giving a null depth $\eta_s/\eta_{p_{peak}}$ of $6.73 \times 10^{-4}$. c) The DM solution found by iEFC targeting only the LP 21b port. The RMS of the control values is 0.0064, corresponding to approximately 22 nm of stroke, five times smaller than the RMS of the control values for the solution targeting all four ports.}
\end{center}
\end{figure*}

Figure \ref{fig:lp21b_only} shows an example iEFC run targeting just the LP 21b port (i.e. the corresponding control matrix $C$ is calculated using a response matrix $S$ that includes only measurements from the LP 21b port). A probe amplitude of 0.02 is used, and a final $\eta_s$ of $7.42\times 10^{-5}$ is achieved, significantly deeper than reached during the experiment targeting all four ports at once. For this experiment, the final $\eta_s$ is extracted from camera photometry and converted to a normalized coupling value based on the calibrated initial measurement. This is because we observed the null degrade significantly in the time it took to measure a coupling map with the photodiode (approximately 30 minutes). This drift in the null also limits our ability to push to deeper nulls by relinearizing the response matrix, since the null also significantly degrades during the calibration process. A model-based algorithm \cite{giveon_efc} or an algorithm that recalibrates on-the-fly may be able to avoid this problem.

However, the peak planet coupling $\eta_{p_{peak}}$ is not significantly impacted by these drifts, so we measure it using the photodiode, obtaining $\eta_{p_{peak}} = 0.110$. The retained planet throughput shows that the iEFC solution is less aggressive when targeting only one port, and ultimately achieves a null depth $\eta_s/\eta_{p_{peak}}$ of $6.73 \times 10^{-4}$. The S/N ratio cross-sections before and after iEFC are presented in Figure \ref{fig:lp21b_only}b, showing that the limitation observed in Fig. \ref{fig:all_ports_run} is not due to this port in particular, but is a consequence of targeting all four ports at once, a phenomenon discussed further in Section \ref{sec:disc}.

\section{Discussion} \label{sec:disc}
We conducted several simulations to further explore and corroborate the behavior observed on the testbed, elucidating some notable properties of performing iEFC with a PLN. Simulations were conducted with both a realistic lantern model (using the measured modes of the real lantern) as well as with a `perfect' lantern model with ideal LP modes. The first set of simulations and their key takeaways are as follows:

\begin{enumerate}
    \item Simulating iEFC through a realistic lantern model corroborates the behavior observed on the testbed, where the LP 11ab and LP 21a ports improve at the expense of the LP 21b port. However, both LP 11ab ports can now reach an $\eta_s$ of $2 \times 10^{-5}$ simultaneously (even deeper if traded-off against each other by relinearizing at different times), likely because there is no instability or noise in the simulation. Meanwhile, the LP 21a port bottoms out at a shallower $\eta_s \approx 2 \times 10^{-3}$, perhaps a tradeoff related to the deeper LP 11 nulls.
    \item Simulating iEFC through a perfect lantern model with static phase-only aberrations in the pupil shows all four ports reaching extremely deep contrasts ($\eta_s \sim 10^{-9}$ to $10^{-12}$ in the absence of other noise, depending on the exact aberration). The static phase aberration in this case has a peak-to-valley (PTV) value of 0.2 radians.
    \item Simulating iEFC through a perfect lantern model with static phase and amplitude aberrations (0.1 radians PTV each) in the pupil shows that the ports are now limited to shallower nulls ($\eta_s \sim 10^{-5}$ to $10^{-7}$ in the absence of other noise, with an observed tradeoff amongst the ports that depends on the exact aberration). 
\end{enumerate}

Back-propagating the measured modes and the ideal LP modes into the pupil plane and taking their difference shows that the cross-coupling of the lantern from manufacturing imperfections appears as both phase and amplitude deviations in the pupil plane. These simulations thus suggest that the cross-coupling manifesting partially as amplitude aberration in the pupil plane is what prevents all four ports from being nulled.

However, the typical coronagraphic solution of adding a second DM to correct for amplitude errors is not a necessary or even particularly effective solution in this case. The observed behavior does not seem to be a fundamental limitation of physics, but rather a unique property of mode-sorters such as the PLN and how iEFC senses the electric field through the ports. For example, simulating model-based EFC with a realistic lantern model given perfect knowledge of the electric field shows that one DM is, in fact, capable of compensating for cross-coupling down to $\eta_s < 10^{-11}$ and beyond. Therefore, one DM has enough \textit{control} authority to compensate for both phase and amplitude errors in the pupil --- e.g. by manipulating phase to balance out amplitude asymmetries such that the overlap integral is still zero.

The limitation, therefore, lies with the sensing: as mentioned in Section \ref{sec:four_ports}, an SVD of the iEFC response matrix $S$ shows that the most dominant modes are approximately Coma X\&Y (primarily impacting the LP 11 ports), followed by $0^{\circ}$ and $45^{\circ}$ Astigmatism (primarily impacting the LP 21 ports), followed by weaker irregular modes. Fully compensating for amplitude asymmetries would require higher spatial frequency modes --- as is observed in the DM solution obtained with EFC given perfect knowledge of the electric field; however, these modes are not well-sensed by the current pairwise probing approach, which mainly captures the effect of the most dominant low order modes.

While improvements in the lantern manufacturing process may reduce cross-coupling to the degree that this is no longer a problem at the contrasts required for ground-based astronomy, these sensing considerations will likely remain relevant at the higher contrasts required for space telescope instrumentation. Therefore, future work includes exploring alternative methods for sensing the electric field, and building response matrices in a way that addresses the unique physical characteristics of mode-sorters, where the outputs are not pixels directly sensing the local intensity but rather ports whose coupling is determined by an overlap integral. This work would be applicable to the PLN and as well as other designs such as the Single-mode Complex Amplitude Refinement (SCAR) coronagraph \cite{por_scar, haffert_scar}. 

Meanwhile, the experiment targeting the LP 21b port alone is limited by a rapid drift in the null. This behavior differs from those of the LP 11 nulls when targeting all four ports at once, which exhibit large fluctuations but very little degradation over time. Based on theoretical sensitivities to aberrations \cite{xin_2022}, we believe this is because the LP 11 nulls are primarily affected by zero-mean tip-tilt fluctuations, while the LP 21 nulls are primarily affected by higher order modes (as can be inferred from the DM solution in Fig. \ref{fig:lp21b_only}c) that drift over time.

The combined behavior of instrument sensitivity and the aberrations' spatial and temporal characteristics, the layout of the adaptive optics (AO) system, and whether the planet location is already known --- these factors all have implications for the use of iEFC for on-sky observations. For example, if the planet location is already known, then it is more advantageous to target one or two ports, maximizing sensitivity at the planet's location. However, for a survey, or for a planet whose location is not well-constrained, it would be necessary to target either both LP 11 ports, both LP 21 ports, or all four ports at once in order to obtain more coverage of the sky.

The choice of which ports to target also depends on whether the AO system is set up to perform iEFC in real time during the observation. Some systems, such as the Keck Planet Imager and Characterizer (KPIC) \cite{delorme_kpic}, have a dedicated second-stage DM that is not seen by the primary AO wavefront sensor. This means that the DM can be changed during the course of the observation without impacting the AO loop, as is done by KPIC to perform speckle nulling through the single-mode fiber \cite{xin_sn} --- a control architecture would allow the iEFC loop to run independently on sky. However, many instruments do not have this extra DM, so any changes to the DM made by the iEFC loop have to be accounted for with an `offset' to the primary AO wavefront sensor. This process adds significant complexity, and though it has been demonstrated on sky with a Shack-Hartmann wavefront sensor \cite{potier_2022}, it is more difficult with Pyramid wavefront sensors, with which it has not yet been reliably implemented on-sky. Although any combination of ports could be targeted on sky using systems with a second-stage DM, since iEFC can be run in real time, for instruments without a second-stage DM, a more viable strategy would be to run iEFC during daytime calibrations, then observe in open loop. This method relies on the null being relatively stable, or primarily sensitive to aberrations that fluctuate about a zero mean in time, which can inform the choice of ports used (i.e. the LP 11 ports on the PoRT testbed, though this would depend on environmental conditions). Although there is no primary AO loop for space-based telescopes, similar interactions may arise with upstream low-order wavefront sensors \cite{pourcelot_2023}, and additional effort is needed to ensure that both the low-order and focal-plane loops can run concurrently.

An additional consideration for ground-based telescopes is whether to observe using pupil-tracking or field-tracking mode. Running iEFC actively on-sky would benefit from a more stable pupil-to-instrument relationship, but would require a more complicated data analysis approach to fit the rotationally-varying planet signal (Goyal et al. in prep). Depending on various factors (such as the speed of sky rotation, the rate of quasi-static speckle evolution, and the planet brightness and separation), the rotating planet signal (after post-processing with a modified version of Angular Differential Imaging \cite{marois_adi}) may or may not provide an overall gain in detection sensitivity compared to the static planet signal from field-tracking mode. Meanwhile, if iEFC is used to calibrate the PLN during the day to compensate for static bench and lantern imperfections without seeing the pupil, then either mode can be used depending on which provides the best post-processed performance, which is currently an open question. Additional work and empirical experience is needed to identify the best approach for scientific observations, including considerations of robustness and ease of use.

Additionally, this demonstration uses a laser due to the limitations of the photometric backend, but broadband nulls would be necessary for spectroscopic applications. Laboratory results presented in Ref. \citenum{xin_2024_pln_lab} indicate that the nulls of the PLN are naturally broadband at the $\eta_s \sim 10^{-3}$ level, and multi-wavelength control scales more advantageously for the PLN than for coronagraphs, as there are fewer degrees of freedom per wavelength that must be controlled by the DM. However, future work is needed to either demonstrate that monochromatic control can achieve a null that is sufficiently spectrally wide enough for science, or to demonstrate full multi-wavelength control. 

\section{Conclusion}

In this work, we showed that implicit electric field conjugation, a focal-plane wavefront control method developed for coronagraphs, can also be used to improve the null depths of the PLN. We find that there is a tradeoff amongst the ports, where deepening some of the nulls can occur at the expense of the others, but overall, the null depths can be improved by one to two orders of magnitude for up to three of the four ports at once. We conduct simulations that corroborate and provide additional insight into the behavior observed on the testbed, as well as reveal an electric field sensing problem for mode-sorters that will be explored in future work. Future work also includes chromatically dispersing the outputs of the PLN, performing iEFC in broadband light, and characterizing the null depths achievable across the 10-20\% bandwidths applicable to high-resolution spectroscopy science. Although the symmetries of the mode-selective photonic lantern prohibit it from being used as a linear wavefront sensor \cite{Lin_PLWFS1}, it may be possible to design a hybrid lantern with some asymmetric ports and some symmetric ports, allowing for nulling and wavefront sensing simultaneously without the need for DM probes. Additionally, preparations for on-sky demonstrations of the PLN are also currently underway.

\subsection*{Disclosures}
The authors have no relevant financial interests in the manuscript and no other potential conflicts of interest to disclose.

\subsection* {Code, Data, and Materials Availability}
The code, data, and simulations used in this work can be found at \url{https://github.com/yinzi-xin/pln_iefc_jatis}, or with the DOI: 10.5281/zenodo.15115353.

\subsection* {Acknowledgments}
We thank the anonymous reviewers for their valuable feedback and suggestions for improving this manuscript. Y.X. acknowledges support from the National Science Foundation Graduate Research Fellowship under Grant No. 1122374. Additional effort has been supported by the National Science Foundation under Grant Nos. 2109231,  2109232, 2308360, and 2308361. This research was carried out in part at the California Institute of Technology and the Jet Propulsion Laboratory under a contract with the National Aeronautics and Space Administration (NASA). This research made use of hcipy \cite{por2018hcipy}; Astropy \cite{astropy:2013,astropy:2018,astropy:2022}; NumPy \cite{harris2020array}; SciPy \cite{2020SciPy-NMeth}; and Matplotlib \cite{Hunter:2007_matplotlib}. An earlier version of this work was presented in the Proceedings of SPIE in Ref. \citenum{xin_spie_2024}.


\bibliography{report}   
\bibliographystyle{spiejour}   


\vspace{2ex}\noindent\textbf{Yinzi Xin} is a graduate student at Caltech, where she works in the Exoplanet Technology Lab under the guidance of her advisor, Dimitri Mawet. Her research interests lie in the field of high contrast imaging instrumentation and data analysis for exoplanets. She is interested in wavefront sensing and control, coronagraphy, and the development of new instrument concepts. 

\vspace{1ex}
\noindent Biographies and photographs of the other authors are not available.

\end{spacing}
\end{document}